\documentclass[twocolumn,showpacs,preprintnumbers,amsmath,amssymb]{revtex4}
\usepackage{graphicx}% Include figure files
\usepackage{dcolumn}% Align table columns on decimal point
\usepackage{bm}% bold math

\begin{document}

%%\preprint{APS}

\title{ EIGENVALUE DISTRIBUTIONS FOR A CLASS OF COVARIANCE MATRICES\\
WITH APPLICATIONS TO BIENENSTOCK-COOPER-MUNRO NEURONS UNDER NOISY
CONDITIONS}
\author{Armando Bazzani}
\email{armando.bazzani@bo.infn.it}
\author{Gastone C Castellani}
\affiliation{Department of Physics and National Institute of Nuclear Physics, University of Bologna and\\
Institute for Brain and Neural Systems, Brown University}
%%\altaffiliation[Also at ]{Physics Department, XYZ University.}%Lines break automatically or can be forced with \\
\author{Leon N Cooper}
\affiliation{Department of Physics, Brown University, Providence,
RI 02912 and\\
Institute for Brain and Neural Systems, Brown University}

%%\homepage{http://www.Second.institution.edu/~Charlie.Author}

\date{\today}% It is always \today, today,

\begin{abstract}
We analyze the effects of noise correlations in the input to, or
among, BCM neurons using the Wigner semicircular law to construct
random, positive-definite symmetric correlation matrices and
compute their eigenvalue distributions. In the finite dimensional
case, we compare our analytic results with numerical simulations and
show the effects of correlations on the lifetimes of synaptic
strengths in various visual environments. These correlations can be
due either to correlations in the noise from the input LGN neurons,
or correlations in the variability of lateral connections in a
network of neurons. In particular, we find that for fixed
dimensionality, a large noise variance can give rise to long
lifetimes of synaptic strengths. This may be of physiological
significance.

\end{abstract}

%Use showkeys class option if keyword
\maketitle

\section{Introduction}

Receptive field changes in visual cortex, observed in the early
period of an animal's postnatal development, are thought to depend
on synaptic plasicity\cite{bear99,seng02}; the detailed dynamics of
such receptive field modifiability has been used to infer the
precise form of this plasticity\cite{blais98,cooper04}. In a classic
paradigm, called monocular deprivation, vision through one eye is
deprived in early development. In this case cells in visual cortex
tend to disconnect from deprived eye\cite{wiesel62}. Experiments have shown that
if monocular deprivation is produced by monocular lid closure then a
rapid loss of response to the deprived eye occurs, while if the
retina is inactivated by a Tetrodotoxin (TTX) injection,
significantly less loss is observed\cite{ritten99,frenk04}.
These results are consistent with the form of synaptic plasticity proposed by Bienenstock, Cooper
and Munro (BCM)\cite{bienen82}.
The BCM theory was originally proposed to describe plasticity processes
in visual cortex as observed by Hubel and Wiesel \cite{wiesel62}. One of the main postulates of this theory is the existence
of a critical threshold (the sliding threshold) that depends on past neuronal history in a non-linear way.
This nonlinearity is necessary to ensure stability of the synaptic weights. The main predictions of the BCM theory have been
confirmed in hippocampal slices and visual cortex and recently in in vivo inhibitory avoidance learning experiments\cite{bear2006}.
The extension of these results to other brain areas, and ultimately to the whole brain, is not confirmed but is under active study.
One motivation for  this research is that a  proposed  biophysical mechanism  for the BCM rule is based on calcium influx through
NMDA receptors and phosphorylation state of AMPA receptors and that both receptors are widely distributed within the brain\cite{cast01}.
This  biophysical mechanism is at least partly shared, by the plasticity rule STDP (Spike-timing-dependent plasticity)\cite{abbott2000synaptic} that
describes the synaptic functional change on the basis of action potentials timing in connected neurons.
The main difference between STDP and BCM is that BCM  is an average time rule and thus does not include microscopic temporal structures
(i.e. it works with rates not spikes)\cite{gerstner1996neuronal,gerstner2002mathematical}.
 A further analysis that considers the relation between BCM and
STDP rules will be considered in a future work.
\par\noindent
The standard theoretical analysis of monocular deprivation experimental results,
according to BCM (see for example \cite{ cooper04}) relies on the, seemingly reasonable, assumption that in the
inactivated situation, activity in the lateral geniculate nucleus
(LGN), which is the cortical input, is reduced compared to the lid
closure case; as a consequence there is a less rapid loss of
response to the deprived eye. This assumption has been questioned by
new experimental results.\cite{lind09}\par In this recent study the
activity of neurons in LGN has been measured during normal vision,
when the eye-lid of the experimental animals was sutured and when
TTX was injected into the lid sutured eye. The recordings were made
on awake animals while they watched movie clips and sinusoidal
gratings. A surprising result of these experiments is that
inactivation does not reduce mean activity in LGN compared to lid
suture; however inactivation produced an increase in correlations
between different cells within the LGN. Previous experimental
results in ferret LGN\cite{weli99,ohsh06}, and further results in mouse LGN\cite{lind09} indicate
that the activity of nearby cells in LGN are correlated, that this
activity falls off as a function of the distance between the
receptive fields of the cells, and that these correlations exist
even in the absence of retinal activity.\par\noindent A recent
paper\cite{blais09} has examined the impact of input correlations
during deprivation experiments and has shown that correlations in
LGN noise can significantly slow down the loss of response to the
deprived eye for BCM neurons\cite{bienen82}, in agreement with experiments.
This paper also examines the effect of such correlations on a class of
PCA type learning rules. Thus
correlated LGN noise theoretically leads to persistence of synaptic strength for a
period that depends on the level of noise correlation. As a
consequence, noise at neuronal level could play a fundamental role
in synaptic plasticity.  The effect of white noise
on BCM has been previously studied\cite{bazz03}. In this paper we show that noise
correlations in the input or among BCM neurons can be studied by using
the eigenvalue distributions of random positive-definite
symmetric matrices. In a simple but generic case, we explicitly
compute the distribution by using the Wigner semicircular law\cite{wig58,wig67},
pointing out the role of correlations and the existence of a
thermodynamic limit. In the finite dimensional case, the analytic
results are compared with numerical simulations with applications to
real conditions. We also discuss a transition in the eigenvalue
distribution when the noise level is increased. This phenomenon
implies the existence of states with very long lifetimes; these
could have physiological significance in the development of neuronal
systems.

\section{BCM neuron in monocular deprivation}

\begin{figure}
\includegraphics[width=8 truecm]{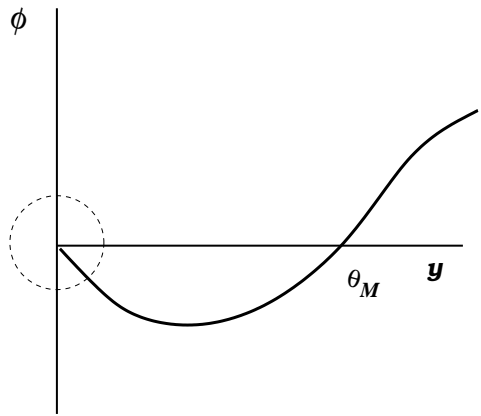}\\ % Here is how to import EPS art
\caption{\label{figure0} The BCM function $\phi(y)$; when
$y$ is close to zero, we can linearize the system (\ref{1.1}) by approximating $\phi\simeq - \epsilon y$.}
\end{figure}

\begin{figure}
\includegraphics[width=8 truecm]{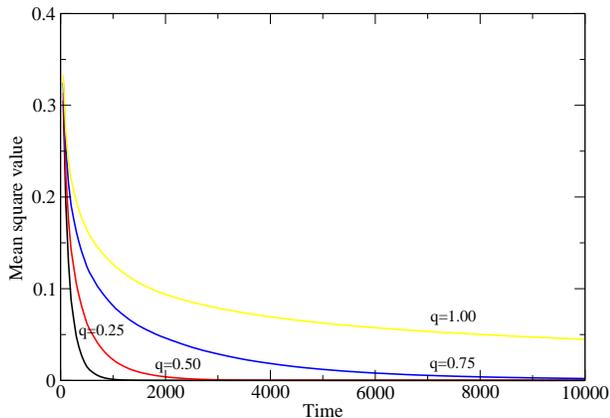}\\ % Here is how to import EPS art
\caption{\label{bcmcorr} Evolution of the norm of BCM synaptic weights when the input is the correlated
noise defined by eq. (\ref{1.corr}). We consider increasing values for $q$: $q=0.25,0.50,0.75,1.00$ for $N=100$. The
curves give the evolution of the mean square value of the synaptic weights $\mathbf{w}$ when the input noise correlation increases keeping
the noise variance fixed (time is in arbitrary units). We note the existence of non zero long time persistent states
when the correlation is high, whereas for low correlation value the system tends to the equilibrium state $\mathbf{w}=0$.}
\end{figure}

We briefly review the behavior of a Bienenstock, Cooper and Munro (BCM) neuron\cite{bienen82} in monocular deprivation.
Let $\mathbf{w}$ be the synaptic weights and $\mathbf{x}$ the input signals received by the synapses,
the BCM synaptic modification rule has the form
\begin{equation}
\mathbf{\dot w}=\mathbf{x}\phi(y,\theta_m) \label{1.1}
\end{equation}
where the modification function $\phi(y,\theta_m)$ depends on the neuron
activity level $y\propto\mathbf{x}\cdot\mathbf{w}$ (it is assumed a linear proportionality between the input $\mathbf{x}$
and the output $y$) and on a moving threshold $\theta_m$, which is a super-linear function of the cell activity history
(in a stationary situation $\theta_m$ can be related to the time averaged value $<y^k>$ where $k>1$ of a non-linear
moment of the neuron activity distribution)\cite{cooper04}. The modification function $\phi$ is a non-linear function
of the postsynaptic activity $y$
which has two zero crossings, one at  $y$=0 and the other at  $y=\theta_m$ (see fig. \ref{figure0}).
When the neuron activity is above the threshold $\theta_m$ we have LTP, whereas LTD appears when
the activity is below the threshold. The nonlinear dependence of the threshold on
neuron activity solves the stability problem of Hebb's learning rule,
preventing a dynamical divergence of synaptic weights ($y=\theta_m$ is an attractive fixed point
for a stationary input)\cite{cooper04}.
In the simplest form the function $\phi$ is a quadratic function $\phi(y)=y^2-y\theta_m$ and the
dynamic threshold $\theta_m$ is the time-averaged $<y^2>$ of the second moment of the neuron activity, which can be replaced
by the expectation value over the input probability space $E(y^2)$ under the slow-learning assumption.
During MD regime,
the input signal $\mathbf{x}$ to the closed eye can be represented by
a stochastic process with zero mean value and small variance ($<\mathbf{x}^2>\ll 1$).
One can numerically study the effect of noise correlation by integrating the system (\ref{1.1}). Let
us introduce the input noise
\begin{equation}
\mathbf{x}(t)=A\xi(t)
\label{1.corr}
\end{equation}
where $\xi(t)$ is a stochastic process in $\mathbb{R^N}$ defined by i.i.d. random variables with zero mean value and normalized
variance. We consider a symmetric matrix $A$ of the form
\begin{equation}
A={1\over \sqrt{1+q^2}}\begin{pmatrix} 1&{q\over \sqrt{N}} a_{12}&...&{q\over \sqrt{N}} a_{1N}\\ {q\over \sqrt{N}} a_{12}&1&...&{q\over \sqrt{N}} a_{2N}
\\ ..&..&...&..\\
{q\over \sqrt{N}} a_{1N}&{q\over \sqrt{N}} a_{2N}&...&1 \end{pmatrix}\qquad  N\gg 1\label{1.noise}
\end{equation}
where the coefficients $a_{ij}$ are independent realizations of a normalized random variable with zero mean value.
The covariance matrix of the noise $\mathbf{x}(t)$ is
$$
A A^T\simeq
\begin{pmatrix}
1&{2q\over \sqrt{N}(1+q^2)}a_{12}&...&{2q\over \sqrt{N}(1+q^2)}a_{1N}\\
{2q\over \sqrt{N}(1+q^2)}a_{12}&1&...&{2q\over \sqrt{N}(1+q^2)}a_{2N}\\
\\ ..&..&...&..\\
{2q\over \sqrt{N}(1+q^2)}a_{1N}&{2q\over \sqrt{N}(1+q^2)}a_{2N}&...&1\\
\end{pmatrix}
$$
Then for a given $N$, we can vary the correlation among the noise components by varying $q\in[0:1]$ keeping fixed the
noise variance. We have simulated the weight dynamics in eq. (\ref{1.1}) by using the modification potential
$\phi(y)=y(y-1)$ where we set the neuron activity $y=(\mathbf{x}\cdot \mathbf{w})/N$ to study the limit $N\gg 1$. In this simple model,
if the activity $y$
is below the threshold $\theta_m=1$ we expect LTD and the synaptic weights $\mathbf{w}$ relax towards 0. To characterize the
relaxation process, we compute the mean square value of the vector $\mathbf{w}(t)$; the results are plotted
in fig. \ref{bcmcorr} for increasing values of $q$. In all cases
the system (\ref{1.1}) tends to a stationary solution. For low correlated noise we recover the equilibrium solution
$\mathbf{w}=0$ as expected, but when the input noise is strongly correlated, the simulations
show the existence of non zero long time persistent states for the synaptic weights. Moreover the numerical simulations
suggest that the persistent states correspond to eigenvectors of the noise covariance matrix with very small eigenvalues.
These results are consistent with experimental observations\cite{lind09} on the
activity of neurons in LGN in the case of monocular deprivation.\par\noindent
A theoretical approach
linearizes the system (\ref{1.1}) around $y\simeq 0$ by considering $\theta_m$ constant. Then eq. (\ref{1.1}) becomes (cfr. fig. \ref{figure0})
\begin{equation}
\mathbf{\dot w}=-\epsilon\mathbf{x}(\mathbf{w}\cdot\mathbf{x})
\label{1.1n}
\end{equation}
so that on average we have
\begin{equation}
\dot w_i=-\epsilon \sum_{j=1}^N C_{ij}w_j
\label{1.2n}
\end{equation}
where $C$ is the noise covariance matrix. In the case of
uncorrelated input ($C_{ij}=n_i^2\delta_{ij}$ where $n_i^2$ is the $i$-component noise variance) equation
(\ref{1.1n}) becomes
$$
\dot w_i=-\epsilon n_i^2 w_i
$$
and $w_i(t)$ tends exponentially to zero with a scale time $\simeq
1/(\epsilon n_i^2)$. In a generic case the system (\ref{1.2n}) can be solved by diagonalizing the
covariance matrix $C$ and expand the synaptic weights $\mathbf{w}$ in the eigenvector base. Then
each component $w_\lambda$ in this base will evolve according to
$$
w_\lambda(t)=w_\lambda(0)e^{-\lambda t}
$$
where $\lambda$ is the corresponding eigenvalue of the covariance matrix. As a consequence when $\lambda \ll 1$ we have
long time non trivial persistent states for the synaptic weights. According to numerical simulations shown
in the fig. \ref{bcmcorr} this is the case for correlated noisy input.
In order to perform explicit calculations we assume that the covariance matrix is a random
positive definite symmetric matrix of the form
\begin{equation}
C_{kl}=n^2(C^0_{kl}+m V_{kl}+O(m^2)) \label{1.11}
\end{equation}
where
\begin{equation}
C^0=\begin{pmatrix}1&q&...&q\\ q&1&...&q\\ ..&..&...&..\\
q&q&...&1 \end{pmatrix}\qquad q\le 1 \label{2}
\end{equation}
and $V_{kl}$ is a symmetric random matrix with zero mean value,
normalized variance and i.i.d. elements. The parameter $q\in [0,1]$
determines the average correlation among the inputs and the parameter
$m^2$ is the variance of fluctuations of the covariance matrix. The matrix (\ref{2}) has the
following eigenvalues
\begin{equation}
\lambda_i=\begin{cases}1+(N-1)q\qquad i=1\\ 1-q\qquad i=2,..,N\\
\end{cases} \label{5}
\end{equation}
Therefore when $q\to 1$ (i.e. increase of the average correlation) the eigenvalues $\lambda_{2,..,N}$ tend to
0 linearly in $(1-q)$ and the correlation matrix $C^0$ is
degenerate.  When $m=0$, the effect of correlations is to
increase the decay time of $w^{(k)}$ $\propto
1/(1-q)$. In the next section we analyze the effect of
fluctuations $V_{kl}$ on the eigenvalue distributions of the matrix
(\ref{1.11}) in the thermodynamics limit $N\to\infty$.

\section{ Eigenvalue distributions}

We consider the eigenvalue distributions of the matrix
$C_{kl}$ (see eq. (\ref{1.11})) using the unperturbed covariance
matrix (\ref{2}). Since $C_{kl}$ is a symmetric positive definite
matrix we can introduce the following representation
\begin{equation}
C=(\sqrt{C^0}+m W)^2=C^0+m(\sqrt{C^0}W+W\sqrt{C^0})+m^2 W^2
\label{3}
\end{equation}
where $\sqrt{C_0}$ is the symmetric square root of the matrix
(\ref{2}) and and $W$ is a symmetric random matrix with i.i.d.
elements and zero mean value. By direct calculation we obtain the
relation between $V$ and $W$
$$
V=\sqrt{C^0}W+W\sqrt{C^0},
$$
Then $W$ is uniquely determined as a function of $V$ and the
expectation value of the covariance matrix
$$
<C>=C^0+Nm^2I_N
$$
In order to control the noise variance in the thermodynamic limit, $N\to \infty$, we introduce the scaling
parameter $m_0$
\begin{equation}
m=m_0\sqrt{1-q\over N} \label{4}
\end{equation}
This is consistent with the noise definition in the numerical simulations (cfr. eq. (\ref{1.noise})).
The $\sqrt{1-q}$ dependence allows us to control the limit $q\to 1$
when the unperturbed matrix $C^0$ is degenerate. In real systems both $q$ and $N$ are fixed by physical conditions and $m_0$ is
proportional to the fluctuations of the covariance matrix. In order to study
the eigenvalue distribution of the covariance matrix $C$ we consider
the case when $W$ is a Wigner matrix\cite{mehta04}. Then Wigner's
theorem\cite{wig58,wig67} on the eigenvalue distribution can be stated as:\vskip 1. truecm
\par\noindent {\sl Let $W_{ij}$ be a symmetric $N\times N$ random
matrix with i.i.d. elements whose common distribution has zero mean
value, normalized variance and is even, then the probability
distribution of the eigenvalues in the interval
$[\sqrt{N}\alpha,\sqrt{N}\beta]$
$$
P_N(\sqrt{N}\alpha<\mu<\sqrt{N}\beta)\qquad -2<\alpha<\beta<2
$$
is in the thermodynamics limit
$$
\lim_{N\to\infty}P_N(\sqrt{N}\alpha<\mu<\sqrt{N}\beta)=\int_\alpha^\beta
{(4-\mu^2)^{1/2}\over 2\pi}d\lambda
$$
}
\par\noindent
To treat the problem of characterizing the eigenvalue
distribution function when the covariance matrix $C$ has the form
(\ref{3}), we first perform the orthogonal transformation $O$ that
diagonalizes the unperturbed matrix (\ref{2}). We then restrict our
analysis to the invariant subspace that corresponds to the
degenerate eigenvalue $1-q$. In the thermodynamic limit the
eigenvalue $1+(N-1)q$ is singular and decouples the corresponding
invariant subspace. The covariance matrix (\ref{3}) has the form
\begin{equation}
C'=OCO^T=\left (\sqrt{1-q}I_N+m_0\sqrt{1-q\over N}\hat W\right
)^2\qquad N\gg 1 \label{6}
\end{equation}
where $\hat W=OWO^T$. The eigenvalue equation is
\begin{equation}
{\rm det}\left [\lambda -\left (\sqrt{1-q}I_N+m_0\sqrt{1-q\over
N}\hat W\right )^2\right ]=0 \label{7}
\end{equation}
This is equivalent to the condition
\begin{equation}
{\rm det}[(\sqrt{N} \mu I_N-\hat W)]=0 \label{8}
\end{equation}
where we have the relation
\begin{equation}
\lambda=(1+m_0\mu)^2(1-q) \label{9}
\end{equation}
and $\mu$ are the eigenvalues of a Wigner matrix scaled by
$\sqrt{N}$. As a consequence we have the following Lemma:\vskip 1.
truecm\par\noindent {\sl Any eigenvalue $\lambda$ of the matrix
(\ref{6}) can be written in the form (\ref{9}) where $\sqrt{N}\mu$
is an eigenvalue of the Wigner matrix $W$ so that $\mu\in
(-2,2)$}\par\noindent
The relation (\ref{9}) leads to the existence
of two regions in the $\lambda$ spectrum according to the sign of
$m_0\mu+1$. When $\mu>-1/m_0$ we invert the relation (\ref{9}) by
taking the positive branch of the square root
\begin{equation}
\mu={\sqrt{\lambda}-\sqrt{1-q}\over m_0\sqrt{1-q}} \label{10}
\end{equation}
but when $\mu<-1/m_0$ we have to consider the negative branch
\begin{equation}
\mu={-\sqrt{\lambda}-\sqrt{1-q}\over m_0\sqrt{1-q}} \label{11}
\end{equation}
The value $\mu=1/m_0$ is a critical value for the $\lambda$ spectrum
and we expect a singularity in the eigenvalue distribution. The second case occurs only if $m_0>1/2$, so that there
exists a critical threshold in the noise level that induces a
transition in the $\lambda$ distribution. By applying Wigner's
theorem to obtain the distribution density of $\mu$ in the
thermodynamic limit we obtain
$$
\lim_{N\to\infty}\rho(\mu)={(4-\mu^2)^{1/2}\over 2\pi}
$$
so that in the case $m_0<1/2$ the distribution $\rho(\lambda)$
follows from the definition (\ref{10}) according to\cite{march67,pastur96}
\begin{equation}
\rho(\lambda)=\rho(\mu){d\mu\over
d\lambda}={\sqrt{4m_0^2-(\sqrt{\lambda/(1-q)}-1)^2}\over
4\sqrt{(1-q)\lambda}m_0^2\pi} \label{12}
\end{equation}
where
$$
(1-q)(1-2m_0)^2<\lambda<(1-q)(1+2m_0)^2
$$
In the case $m_0>1/2$ we split the $\lambda$ distribution in two
parts: when
$$
(1-q)(2m_0-1)^2<\lambda<(1-q)(1+2m_0)^2
$$
$\rho(\lambda)$ is given by eq. (\ref{12}), whereas when
$$
0<\lambda<(1-q)(2m_0-1)^2
$$
the distribution is
\begin{eqnarray}
\rho(\lambda)&=&{\sqrt{4m_0^2-(\sqrt{\lambda/(1-q)}-1)^2}\over
4\sqrt{(1-q)\lambda}m_0^2\pi}\nonumber \\
&+&{\sqrt{4m_0^2-(\sqrt{\lambda/(1-q)}+1)^2}\over
4\sqrt{(1-q)\lambda}m_0^2\pi} \label{13}
\end{eqnarray}
In the second case the distribution is continuous for
$\lambda>0$ and singular at $\lambda\to 0$. This result leads to the
possibility that the correlation matrix (\ref{3}) has eigenvalues
arbitrarily close to zero and larger than one (corresponding to anticorrelation effects in the noise).
The unperturbed correlation $q$ among the
initial processes $\mathbf{x}(t)$ introduces what is essentially a scaling factor that
changes the existence interval of the $\lambda$ spectrum. The critical value $m_0=1/2$, corresponds to
a critical value of the fluctuation variance $m^2$ (cfr. definition (\ref{4}))
\begin{equation}
m^2_{crit}={1-q\over 4 N} \label{14}
\end{equation}
From a biophysical point of view, the results (\ref{14}) implies that
for sufficiently correlated noise there exist eigenvectors
$w^{(k)}$ of the synaptic weights that are preserved for a very
long time. This might be a possible mechanisms of
maintaining synaptic weights in neuronal systems in
correlated noisy environments.

\section{ Numerical results on finite dimension covariance matrices}

\begin{figure}
\includegraphics[width=8 truecm]{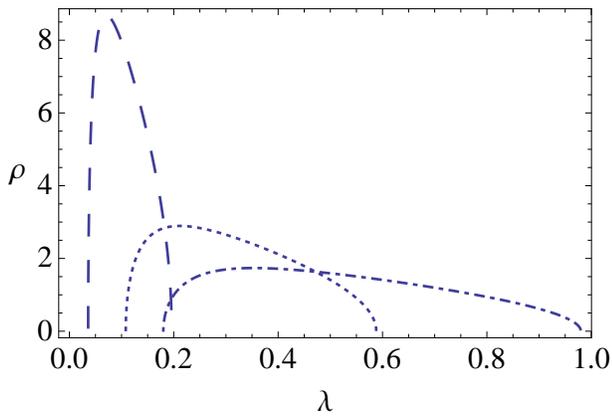}\\ % Here is how to import EPS art
\caption{\label{figure1}Plots of eigenvalue distributions $\rho(\lambda)$
(see definition (\ref{12})) for the correlation matrix (3) using $m_0=.2$ and
$q=.5,.7,.9$ respectively the dashed-dotted, the dotted and the dashed line. As the average correlation $q$
increases the eigenvalue spectrum is squeezed towards the origin.}
\end{figure}

\begin{figure}
\includegraphics[width=6 truecm]{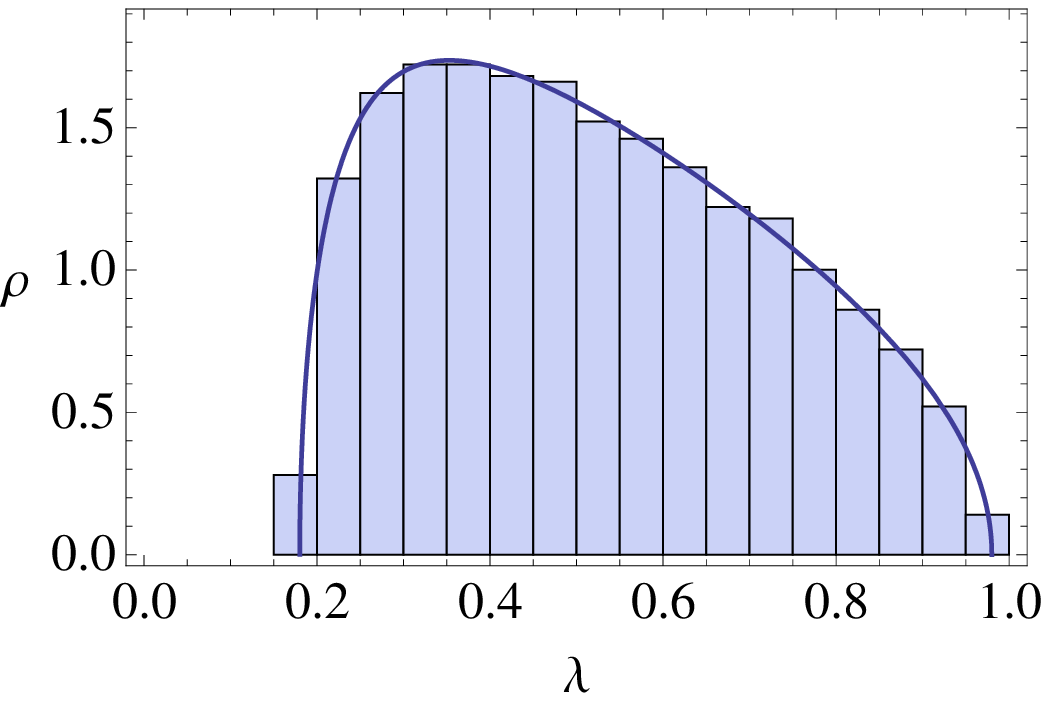}\qquad \includegraphics[width=6 truecm]{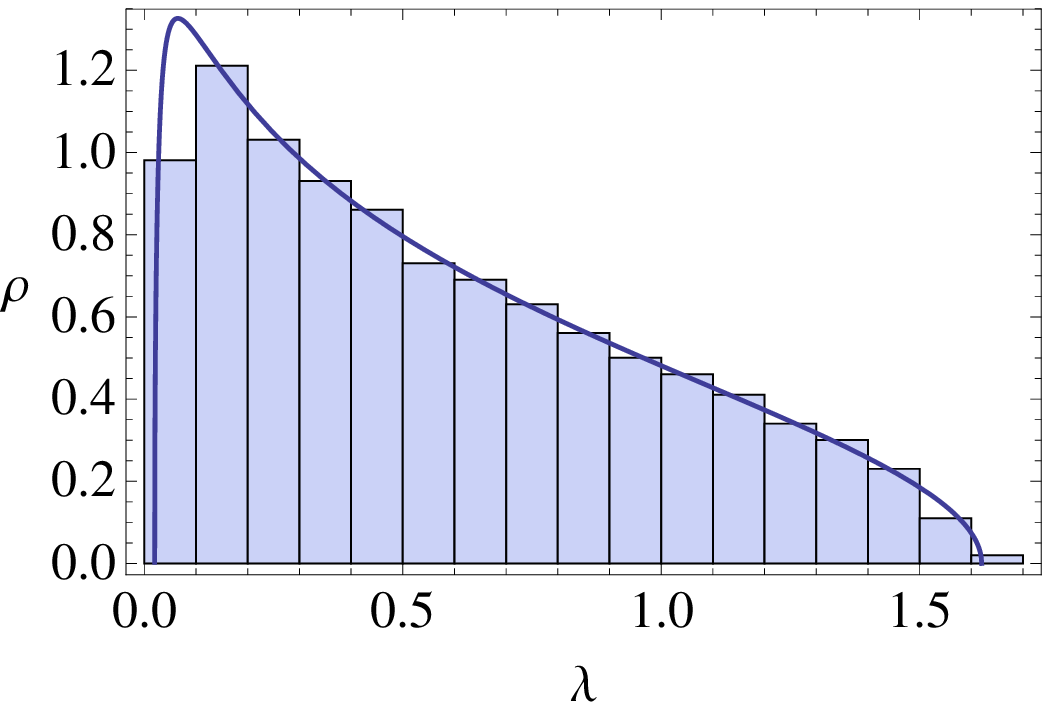}\\
\includegraphics[width=6 truecm]{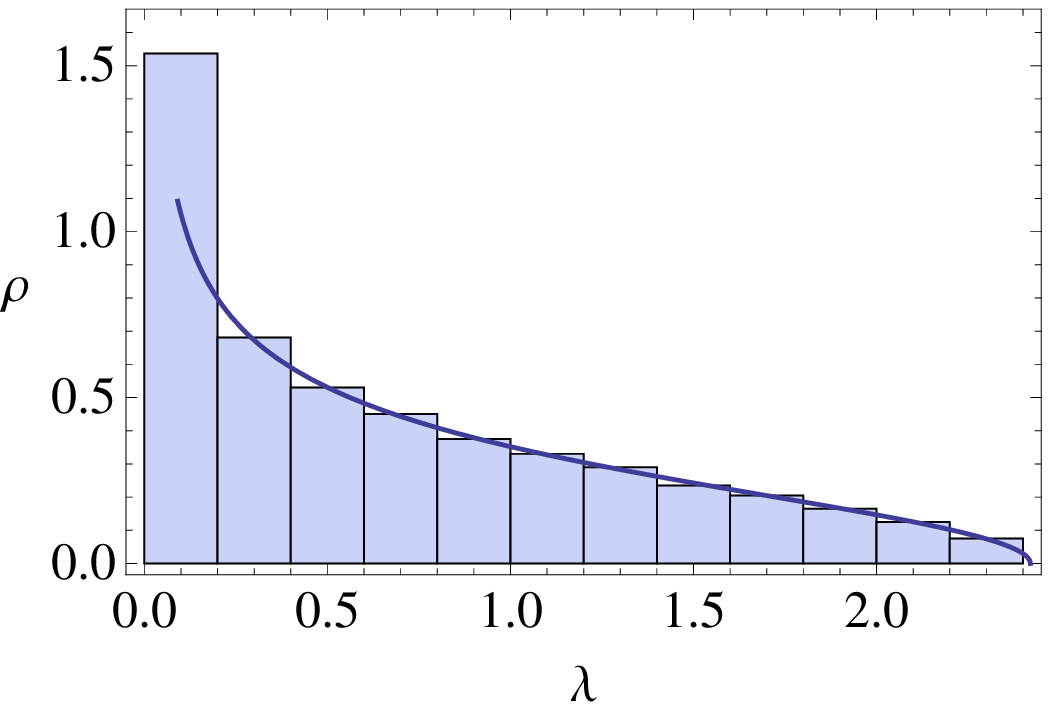}\qquad
\includegraphics[width=6 truecm]{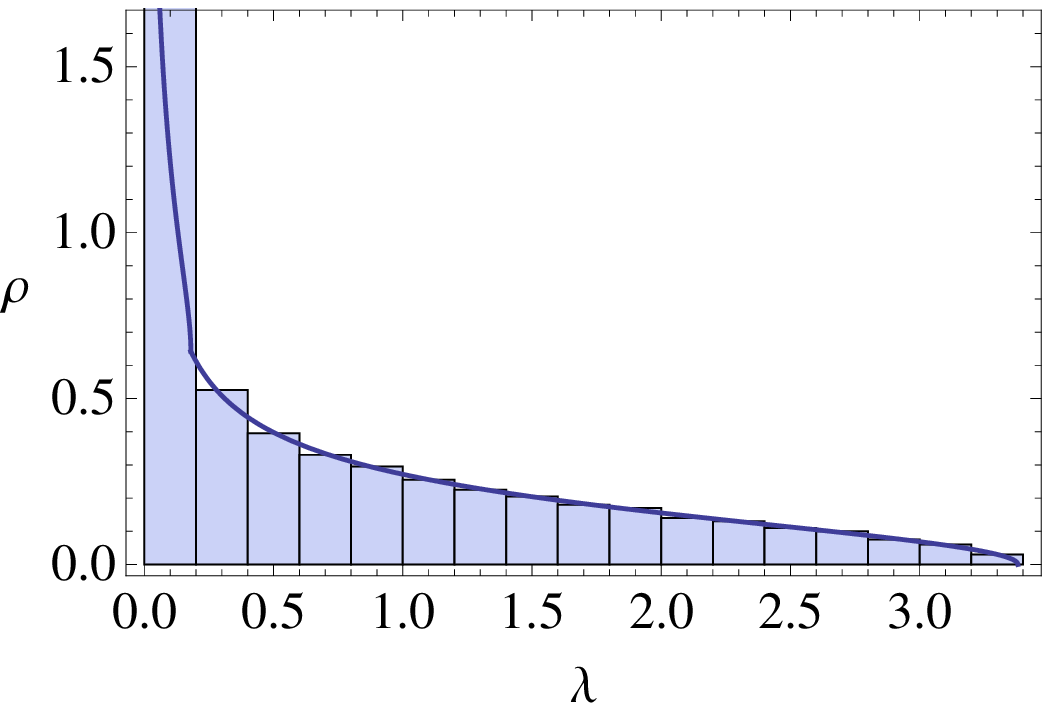}\\
\caption{\label{figure2} Plot of the eigenvalue distribution for the
correlation matrix (3) using $q=.5$ and $m_0=.2,.4,.6,.8$ from left to right and top to
bottom. The effect of the transition at the critical value $m_0=.5$
is clearly seen: the eigenvalues accumulate at the origin. The
continuous line is the theoretical distribution (\ref{12})
whereas the histograms are numerically computed using a random
matrix of order $N=10^3$.}
\end{figure}

To study the effects of finite dimensions on the theoretical results
(\ref{12}) and (\ref{13}) we have performed some numerical
simulations using matrices of dimension $N=10^3$ that might be
realistic for a neural network in LGN. In fig. \ref{figure1} we show
the eigenvalue distribution (\ref{12}) for $m_0=.2$ and different
values of the correlation ($q=.5,.7,.9$).
To see the effect of the transition in the distribution function at
the critical value (\ref{14}), we computed the eigenvalues
distribution in the cases $q=.5$ and $m_0=.4,.6,.8$. The results are
shown in fig. \ref{figure2}

We observe, as in the case of low correlation, the eigenvalue
distributions become larger than in the correlated case and have
values beyond 1 when $m_0$ increases. When $m_0\to 1/2$ the
distribution becomes singular and the distribution peak tends to
zero, so that there exist eigenvectors $w^{(k)}$ with extremely
long decay times in eq. (\ref{1.1n}).

\section{Generalization to a neural network}

For a network of neurons, eq. (\ref{1.1}) can be generalized \cite{castel99} to
\begin{equation}
\mathbf{\dot
w}^{(k)}=\mathbf{x}^{(k)}\phi(y^{(k)},\theta_m^{(k)})\qquad
k=1,...,N \label{1.2}
\end{equation}
where the index $k$ refers to the $k$-th neuron of the network and $N$
is the number of neurons. We then have
\begin{equation}
y^{(k)}=\mathbf{x}^{(k)}\cdot\mathbf{w}^{(k)}+\sum_{l=1}^N
L_{kl}y^{(l)} \label{1.3}
\end{equation}
where we have introduced lateral connections, $L$, among the
neurons. We can have both excitatory or inhibitory
neural connections according to the sign of $L_{kl}$.
In what follows we assume that $L$ is a symmetric matrix
with $\Vert L\Vert\ll 1$ where $\Vert\quad\Vert$ denotes the usual
matrix norm. Therefore the neurons form a bidirectional symmetric
network; we can invert the relation (\ref{1.3}) to obtain
\begin{equation}
y^{(k)}=\sum_{l=1}^N(I-L)^{-1}_{kl}\mathbf{x}^{(l)}\cdot\mathbf{w}^{(l)}
\label{1.4}
\end{equation}
For monocular deprivation eq. (\ref{1.3}) becomes a network generalization of eq. (\ref{1.1n})
\begin{equation}
\mathbf{\dot w}^{(k)}=-\epsilon^{(k)} \mathbf{x}^{(k)}y^{(k)}\qquad
k=1,...,N \label{1.5}
\end{equation}
where $\epsilon^{(k)}$ are suitable positive constants. By
substituting eq. (\ref{1.4}) into (\ref{1.5}), we
get the linear system
\begin{equation}
\mathbf{\dot w}^{(k)}=-\epsilon^{(k)} \mathbf{x}^{(k)}
\sum_{l=1}^N(I-L)^{-1}_{kl}\mathbf{x}^{(l)}\cdot\mathbf{w}^{(l)}
\qquad k=1,...,N \label{1.6}
\end{equation}
In what follows we regard the inter-neuron connections $L$, as not-modifiable. Moreover for
simplicity we reduce external input to one dimension for each
neuron: then both $\mathbf{w}^{(k)}$ and $\mathbf{x}^{(k)}$ are scalar.
In order to study the dynamical properties of the solutions we
average on fast scale variations of the noise; the equations
(\ref{1.6}) become
\begin{equation}
\dot w^{(k)}=-\epsilon^{(k)}\sum_{l=1}^N(I-L)^{-1}_{kl}C_{kl}w^{(l)}
\qquad k=1,...,N \label{1.7}
\end{equation}
where $C_{kl}$ is the noise covariance matrix between the inputs of
the $k$ and $l$ neuron. We consider the connection matrix $L$
as a symmetric random matrix, so that according to our hypotheses
the matrix $(I-L)^{-1}$ is a symmetric positive definite matrix.
Then (\ref{1.7}) can be written in the form
\begin{equation}
\dot w^{(k)}=-\epsilon^{(k)}\sum_{l=1}^N \hat L_{kl}w^{(l)} \qquad
k=1,...,N \label{1.8}
\end{equation}
where
\begin{equation}
\hat L_{kl}=(I-L)^{-1}_{kl}C_{kl} \label{1.10}
\end{equation}
is still a symmetric positive definite matrix. If the input signals
are independent ($C_{kl}=\sigma_k^2\delta_{kl}$), we obtain
$$
\dot w^{(k)}=-\epsilon^{(k)}n_k^2(I-L)^{-1}_{kk}w^{(k)} \qquad
k=1,...,N
$$
and the $w^{(k)}$ exponentially decay towards zero.
Again the interesting case is when the covariance matrix is not
diagonal and the connections $L$ are defined by a random matrix;
from a biological point of view we are modeling a ensemble of
neurons which are stimulated by correlated inputs and have bilateral
connections with random weights. When $\hat L$ has the form
(cfr. eq. (\ref{1.11}))
\begin{equation}
\hat L_{kl}=C^0_{kl}+m V_{kl}+O(m^2) \label{1.11n}
\end{equation}
we can apply the method used for a single neuron to study the
eigenvalue distribution. But in the network case we have a new
biophysical interpretation of the critical value (\ref{14}): if
number of neurons is sufficiently large and/or the fluctuations in
the bilateral neural connections exceed a threshold, the network is
able to develop $w^{(k)}$ eigenvectors with extremely long lifetimes
in presence of a noisy correlated input. This phenomenon
suggests mechanisms for long network lifetimes in
noisy environments. The possibility that correlated noise plays a role in
the dynamics of a neural network has been investigated in \cite{glatt06}
using different models.

\section{Conclusions}

We have analyzed the effects of noise correlations in the input to,
or among, BCM neurons using the Wigner semicircular law to construct
random, positive-definite symmetric correlation matrices and
computing their eigenvalue distributions. In the finite dimensional
case, our analytic results are compared with numerical simulations.
We thus show the effects of correlations on the lifetimes of the
synaptic strengths in various visual environments. In the case of a single neuron
the noise correlations arise from the input LGN neurons, whereas the correlations
arise also from the lateral connections in a neuron network. If
the dimensionality of system is fixed, we show that when the fluctuations
of the covariance matrix exceed a critical threshold synapses with long lifetimes arise. These results may be of physiological
significance and can be tested experimentally.

\newpage %Just because of unusual number of tables stacked at end
\bibliography{correlation}% Produces the bibliography via BibTeX.

\begin{thebibliography}{25}
\expandafter\ifx\csname natexlab\endcsname\relax\def\natexlab#1{#1}\fi
\expandafter\ifx\csname bibnamefont\endcsname\relax
  \def\bibnamefont#1{#1}\fi
\expandafter\ifx\csname bibfnamefont\endcsname\relax
  \def\bibfnamefont#1{#1}\fi
\expandafter\ifx\csname citenamefont\endcsname\relax
  \def\citenamefont#1{#1}\fi
\expandafter\ifx\csname url\endcsname\relax
  \def\url#1{\texttt{#1}}\fi
\expandafter\ifx\csname urlprefix\endcsname\relax\def\urlprefix{URL }\fi
\providecommand{\bibinfo}[2]{#2}
\providecommand{\eprint}[2][]{\url{#2}}

\bibitem[{\citenamefont{Bear and Rittenhouse}(1999)}]{bear99}
\bibinfo{author}{\bibfnamefont{M.~F.} \bibnamefont{Bear}} \bibnamefont{and}
  \bibinfo{author}{\bibfnamefont{C.~D.} \bibnamefont{Rittenhouse}},
  \bibinfo{journal}{J Neurobiol} \textbf{\bibinfo{volume}{41(1)}},
  \bibinfo{pages}{83} (\bibinfo{year}{1999}).

\bibitem[{\citenamefont{Sengpiel and Kind}(2002)}]{seng02}
\bibinfo{author}{\bibfnamefont{F.}~\bibnamefont{Sengpiel}} \bibnamefont{and}
  \bibinfo{author}{\bibfnamefont{P.~C.} \bibnamefont{Kind}},
  \bibinfo{journal}{Curr Biol} \textbf{\bibinfo{volume}{12(23)}},
  \bibinfo{pages}{R818} (\bibinfo{year}{2002}).

\bibitem[{\citenamefont{Blais et~al.}(1998)\citenamefont{Blais, Intrator,
  Shouval, and Cooper}}]{blais98}
\bibinfo{author}{\bibfnamefont{B.~S.} \bibnamefont{Blais}},
  \bibinfo{author}{\bibfnamefont{N.}~\bibnamefont{Intrator}},
  \bibinfo{author}{\bibfnamefont{H.}~\bibnamefont{Shouval}}, \bibnamefont{and}
  \bibinfo{author}{\bibfnamefont{L.~N.} \bibnamefont{Cooper}},
  \bibinfo{journal}{Proceedings of the National Academy of Sciences}
  \textbf{\bibinfo{volume}{10(7)}}, \bibinfo{pages}{1797}
  (\bibinfo{year}{1998}).

\bibitem[{\citenamefont{Cooper et~al.}(2004)\citenamefont{Cooper, Intrator,
  Blais, and Shouval}}]{cooper04}
\bibinfo{author}{\bibfnamefont{L.~N.} \bibnamefont{Cooper}},
  \bibinfo{author}{\bibfnamefont{N.}~\bibnamefont{Intrator}},
  \bibinfo{author}{\bibfnamefont{B.~S.} \bibnamefont{Blais}}, \bibnamefont{and}
  \bibinfo{author}{\bibfnamefont{H.~Z.} \bibnamefont{Shouval}},
  \emph{\bibinfo{title}{Theory of cortical plasticity}}
  (\bibinfo{publisher}{World Scientific New Jersey}, \bibinfo{year}{2004}).

\bibitem[{\citenamefont{Wiesel and Hubel}(1962)}]{wiesel62}
\bibinfo{author}{\bibfnamefont{T.}~\bibnamefont{Wiesel}} \bibnamefont{and}
  \bibinfo{author}{\bibfnamefont{D.}~\bibnamefont{Hubel}},
  \bibinfo{journal}{Journal of Physiology} \textbf{\bibinfo{volume}{180}},
  \bibinfo{pages}{106} (\bibinfo{year}{1962}).

\bibitem[{\citenamefont{Rittenhouse et~al.}(1999)\citenamefont{Rittenhouse,
  Shouval, Paradiso, and Bear}}]{ritten99}
\bibinfo{author}{\bibfnamefont{C.~D.} \bibnamefont{Rittenhouse}},
  \bibinfo{author}{\bibfnamefont{H.~Z.} \bibnamefont{Shouval}},
  \bibinfo{author}{\bibfnamefont{M.~A.} \bibnamefont{Paradiso}},
  \bibnamefont{and} \bibinfo{author}{\bibfnamefont{M.~F.} \bibnamefont{Bear}},
  \bibinfo{journal}{Nature} \textbf{\bibinfo{volume}{397}},
  \bibinfo{pages}{347} (\bibinfo{year}{1999}).

\bibitem[{\citenamefont{Frenkel and Bear}(2004)}]{frenk04}
\bibinfo{author}{\bibfnamefont{M.~Y.} \bibnamefont{Frenkel}} \bibnamefont{and}
  \bibinfo{author}{\bibfnamefont{M.~F.} \bibnamefont{Bear}},
  \bibinfo{journal}{Neuron} \textbf{\bibinfo{volume}{44}}, \bibinfo{pages}{917}
  (\bibinfo{year}{2004}).

\bibitem[{\citenamefont{Bienenstock et~al.}(1982)\citenamefont{Bienenstock,
  Cooper, and Munro}}]{bienen82}
\bibinfo{author}{\bibfnamefont{E.~L.} \bibnamefont{Bienenstock}},
  \bibinfo{author}{\bibfnamefont{L.~N.} \bibnamefont{Cooper}},
  \bibnamefont{and} \bibinfo{author}{\bibfnamefont{P.~W.} \bibnamefont{Munro}},
  \bibinfo{journal}{Journal of Neuroscience} \textbf{\bibinfo{volume}{2}},
  \bibinfo{pages}{32} (\bibinfo{year}{1982}).

\bibitem[{\citenamefont{Whitlockand et~al.}(2006)\citenamefont{Whitlockand,
  Heynen, MG, and Bear}}]{bear2006}
\bibinfo{author}{\bibfnamefont{J.~R.} \bibnamefont{Whitlockand}},
  \bibinfo{author}{\bibfnamefont{A.~J.} \bibnamefont{Heynen}},
  \bibinfo{author}{\bibfnamefont{M.~G.~S.} \bibnamefont{MG}}, \bibnamefont{and}
  \bibinfo{author}{\bibfnamefont{M.~F.} \bibnamefont{Bear}},
  \bibinfo{journal}{Science} \textbf{\bibinfo{volume}{313}},
  \bibinfo{pages}{1093} (\bibinfo{year}{2006}).

\bibitem[{\citenamefont{Castellani et~al.}(2001)\citenamefont{Castellani,
  Quinlan, Cooper, and Shouval}}]{cast01}
\bibinfo{author}{\bibfnamefont{G.~C.} \bibnamefont{Castellani}},
  \bibinfo{author}{\bibfnamefont{E.~M.} \bibnamefont{Quinlan}},
  \bibinfo{author}{\bibfnamefont{L.~N.} \bibnamefont{Cooper}},
  \bibnamefont{and} \bibinfo{author}{\bibfnamefont{H.~Z.}
  \bibnamefont{Shouval}}, \bibinfo{journal}{Proceedings of the National Academy
  of Sciences} \textbf{\bibinfo{volume}{98(22)}}, \bibinfo{pages}{12772}
  (\bibinfo{year}{2001}).

\bibitem[{\citenamefont{Abbott and B.Nelson}(2000)}]{abbott2000synaptic}
\bibinfo{author}{\bibfnamefont{L.~F.} \bibnamefont{Abbott}} \bibnamefont{and}
  \bibinfo{author}{\bibfnamefont{S.}~\bibnamefont{B.Nelson}},
  \bibinfo{journal}{Nature Neuroscience} \textbf{\bibinfo{volume}{3}},
  \bibinfo{pages}{1178} (\bibinfo{year}{2000}).

\bibitem[{\citenamefont{Gerstner et~al.}(1996)\citenamefont{Gerstner, Kempter,
  van Hemmen, and Wagner}}]{gerstner1996neuronal}
\bibinfo{author}{\bibfnamefont{W.}~\bibnamefont{Gerstner}},
  \bibinfo{author}{\bibfnamefont{R.}~\bibnamefont{Kempter}},
  \bibinfo{author}{\bibfnamefont{J.~L.} \bibnamefont{van Hemmen}},
  \bibnamefont{and} \bibinfo{author}{\bibfnamefont{H.}~\bibnamefont{Wagner}},
  \bibinfo{journal}{Nature} \textbf{\bibinfo{volume}{383}}, \bibinfo{pages}{76}
  (\bibinfo{year}{1996}).

\bibitem[{\citenamefont{Gerstner and Kistler}(2002)}]{gerstner2002mathematical}
\bibinfo{author}{\bibfnamefont{W.}~\bibnamefont{Gerstner}} \bibnamefont{and}
  \bibinfo{author}{\bibfnamefont{W.~M.} \bibnamefont{Kistler}},
  \bibinfo{journal}{Biological Cybernetics} \textbf{\bibinfo{volume}{87}},
  \bibinfo{pages}{404} (\bibinfo{year}{2002}).

\bibitem[{\citenamefont{Arnold et~al.}(2009)\citenamefont{Arnold, Heynen,
  Haslinger, and Bear}}]{lind09}
\bibinfo{author}{\bibfnamefont{M.~L.} \bibnamefont{Arnold}},
  \bibinfo{author}{\bibfnamefont{J.}~\bibnamefont{Heynen}},
  \bibinfo{author}{\bibfnamefont{R.~H.} \bibnamefont{Haslinger}},
  \bibnamefont{and} \bibinfo{author}{\bibfnamefont{M.~F.} \bibnamefont{Bear}},
  \bibinfo{journal}{Nature Neuroscience} \textbf{\bibinfo{volume}{12}},
  \bibinfo{pages}{390} (\bibinfo{year}{2009}).

\bibitem[{\citenamefont{Weliky and Katz}(1999)}]{weli99}
\bibinfo{author}{\bibfnamefont{M.}~\bibnamefont{Weliky}} \bibnamefont{and}
  \bibinfo{author}{\bibfnamefont{L.~C.} \bibnamefont{Katz}},
  \bibinfo{journal}{Science} \textbf{\bibinfo{volume}{285}},
  \bibinfo{pages}{599} (\bibinfo{year}{1999}).

\bibitem[{\citenamefont{Ohshiro and Weliky}(2006)}]{ohsh06}
\bibinfo{author}{\bibfnamefont{T.}~\bibnamefont{Ohshiro}} \bibnamefont{and}
  \bibinfo{author}{\bibfnamefont{M.}~\bibnamefont{Weliky}},
  \bibinfo{journal}{Nat Neurosci} \textbf{\bibinfo{volume}{9(12)}},
  \bibinfo{pages}{1541} (\bibinfo{year}{2006}).

\bibitem[{\citenamefont{Blais et~al.}()\citenamefont{Blais, Cooper, and
  Shouval}}]{blais09}
\bibinfo{author}{\bibfnamefont{B.~S.} \bibnamefont{Blais}},
  \bibinfo{author}{\bibfnamefont{L.~N.} \bibnamefont{Cooper}},
  \bibnamefont{and} \bibinfo{author}{\bibfnamefont{H.}~\bibnamefont{Shouval}},
  \emph{\bibinfo{title}{Effect of correlated lgn firing rates on predictions
  for monocular eye closure vs monocular retinal inactivation}},
  \eprint{submitted for publication}.

\bibitem[{\citenamefont{Bazzani et~al.}(2003)\citenamefont{Bazzani, Remondini,
  Intrator, and Castellani}}]{bazz03}
\bibinfo{author}{\bibfnamefont{A.}~\bibnamefont{Bazzani}},
  \bibinfo{author}{\bibfnamefont{D.}~\bibnamefont{Remondini}},
  \bibinfo{author}{\bibfnamefont{N.}~\bibnamefont{Intrator}}, \bibnamefont{and}
  \bibinfo{author}{\bibfnamefont{G.~C.} \bibnamefont{Castellani}},
  \bibinfo{journal}{Neural Computation} \textbf{\bibinfo{volume}{15(7)}},
  \bibinfo{pages}{1621} (\bibinfo{year}{2003}).

\bibitem[{\citenamefont{Wigner}(1958)}]{wig58}
\bibinfo{author}{\bibfnamefont{E.~P.} \bibnamefont{Wigner}},
  \bibinfo{journal}{The Annals of Mathematics} \textbf{\bibinfo{volume}{67-2}},
  \bibinfo{pages}{325} (\bibinfo{year}{1958}).

\bibitem[{\citenamefont{Wigner}(1967)}]{wig67}
\bibinfo{author}{\bibfnamefont{E.~P.} \bibnamefont{Wigner}},
  \bibinfo{journal}{SIAM Review} \textbf{\bibinfo{volume}{9}},
  \bibinfo{pages}{1} (\bibinfo{year}{1967}).

\bibitem[{\citenamefont{Mehta}(2004)}]{mehta04}
\bibinfo{author}{\bibfnamefont{M.~L.} \bibnamefont{Mehta}},
  \emph{\bibinfo{title}{Random Matrices}}, vol. \bibinfo{volume}{142}
  (\bibinfo{publisher}{Elsevier Pure and Applied Mathematics},
  \bibinfo{year}{2004}).

\bibitem[{\citenamefont{Marchenko and Pastur}(1967)}]{march67}
\bibinfo{author}{\bibfnamefont{V.}~\bibnamefont{Marchenko}} \bibnamefont{and}
  \bibinfo{author}{\bibfnamefont{L.}~\bibnamefont{Pastur}},
  \bibinfo{journal}{Mat. Sb.} \textbf{\bibinfo{volume}{72}},
  \bibinfo{pages}{507} (\bibinfo{year}{1967}).

\bibitem[{\citenamefont{Pastur}(1996)}]{pastur96}
\bibinfo{author}{\bibfnamefont{L.~A.} \bibnamefont{Pastur}},
  \bibinfo{journal}{Annales de l'I.H.P. section A}
  \textbf{\bibinfo{volume}{64}}, \bibinfo{pages}{325} (\bibinfo{year}{1996}).

\bibitem[{\citenamefont{Castellani et~al.}(1999)\citenamefont{Castellani,
  Intrator, Shouval, and Cooper}}]{castel99}
\bibinfo{author}{\bibfnamefont{G.~C.} \bibnamefont{Castellani}},
  \bibinfo{author}{\bibfnamefont{N.}~\bibnamefont{Intrator}},
  \bibinfo{author}{\bibfnamefont{H.}~\bibnamefont{Shouval}}, \bibnamefont{and}
  \bibinfo{author}{\bibfnamefont{L.~N.} \bibnamefont{Cooper}},
  \bibinfo{journal}{Networks :Comput. Neur. Syst.}
  \textbf{\bibinfo{volume}{10}}, \bibinfo{pages}{111} (\bibinfo{year}{1999}).

\bibitem[{\citenamefont{Glatt et~al.}(2006)\citenamefont{Glatt, Busch, Kaiser,
  and Zaikin}}]{glatt06}
\bibinfo{author}{\bibfnamefont{E.}~\bibnamefont{Glatt}},
  \bibinfo{author}{\bibfnamefont{H.}~\bibnamefont{Busch}},
  \bibinfo{author}{\bibfnamefont{F.}~\bibnamefont{Kaiser}}, \bibnamefont{and}
  \bibinfo{author}{\bibfnamefont{A.}~\bibnamefont{Zaikin}},
  \bibinfo{journal}{Phys. Rev. E} \textbf{\bibinfo{volume}{73}},
  \bibinfo{pages}{026216} (\bibinfo{year}{2006}).

\end{thebibliography}

\end{document}